\documentclass[aps,prd,reprint,onecolumn,notitlepage,superscriptaddress,longbibliography,nofootinbib,showkeys]{revtex4-2}
\usepackage[T1]{fontenc}
\usepackage{lmodern}
\usepackage{amsmath,amssymb,mathtools}
\usepackage{mathrsfs}
\usepackage{microtype}
\usepackage{needspace}
\usepackage[hidelinks]{hyperref}
\numberwithin{equation}{section}

\allowdisplaybreaks[2]
\hypersetup{pdftitle={Proof of multiloop high-multiplicity integrand reduction coefficients in the 't Hooft-Veltman scheme},pdfauthor={Shengli Yi},pdfsubject={Partial-fraction integrand reduction},pdfkeywords={Feynman integrals, integrand reduction, partial fractions}}

\begin{document}
\title{Proof of multiloop high-multiplicity integrand reduction coefficients in the 't~Hooft--Veltman scheme}
\author{Shengli Yi}
\email{20232321241@m.scnu.edu.cn}
\affiliation{Institute of Quantum Matter, South China Normal University, Guangzhou 510006, China}
\affiliation{School of Physics, South China Normal University, Guangzhou 510006, China}

\begin{abstract}
In the 't~Hooft--Veltman (tHV) scheme, external momenta lie in a fixed integer-dimensional space, and the propagator denominators of high-multiplicity Feynman integrands obey affine-linear relations. Previous work used these relations to construct a partial-fraction reduction with a compact determinantal formula for its coefficients. We present two independent proofs of this formula. The first proof uses recursive partial-fraction decomposition and induction on the number of affine relations. The second adopts the multiple cut method. After clearing the common denominator, setting the complementary denominators to zero isolates one product, and Cramer's rule gives the complete coefficient. We also discuss the behavior of the formula at special kinematics and give an example in which terms with singular coefficients combine to preserve a finite polynomial identity as the denominator-product basis degenerates.
\end{abstract}
\maketitle

\section{Introduction}\label{sec:introduction}

Dimensional regularization treats divergences in perturbative quantum field theory by analytic continuation in the spacetime dimension~\cite{tHooftVeltman1972,BolliniGiambiagi1972,Ashmore1972}. In the 't~Hooft--Veltman (tHV) scheme, loop momenta remain in \(D=4-2\epsilon\) dimensions, while external momenta are kept in a fixed integer-dimensional space~\cite{tHooftVeltman1972}. If the external momenta span a space of dimension \(d_0\), the number of independent loop-dependent scalar products is bounded at fixed loop order. Propagator sets exceeding this bound obey affine-linear relations, which allow partial-fraction reduction under suitable kinematic conditions~\cite{GluzaKajdaKosower2011,Kleiss2012,BargielaYang2024}.

Dimensional constraints have long been used to reduce high-point one-loop integrals~\cite{Melrose1965,vanNeervenVermaseren1984,BinothGuilletHeinrich2000}, including scalar and tensor reductions with four-dimensional external momenta and explicit formulas for five- and six-point tensor integrals~\cite{BinothGuilletHeinrich2000,DennerDittmaier2003,DennerDittmaier2006}. At higher loop order, denominator reducibility has been studied through scalar-product counting, combinations of denominators equal to \(1\), and partial-fraction rules for dependent propagators~\cite{GluzaKajdaKosower2011,Kleiss2012,Pak2012}. The coefficients in these decompositions may be independent of loop momenta or may themselves be polynomials in loop-momentum components, depending on the allowed ansatz~\cite{Kleiss2012}.

For arbitrary loop order in the tHV scheme, Bargie\l a and Yang used affine relations among propagators to construct finite basis topologies and obtained a compact determinantal expression for the partial-fraction coefficients~\cite{BargielaYang2024}. This framework was subsequently applied to two-loop Standard Model integrals~\cite{BargielaYang2025,BargielaEtAl2026}. In the computation of the complete function space for planar two-loop six-particle scattering~\cite{HennEtAl2025}, an eleven-propagator integral representation inspired by this framework was used to speed up the numerical evaluation with AMFlow~\cite{LiuMa2023,HuangLiuMa2026}. A related computation of planar two-loop six-point integrals, which exploits the constraints of four-dimensional external kinematics, was given in Ref.~\cite{AbreuEtAl2025}. Such denominator reductions take place before integration and can simplify the integral families entering integration-by-parts (IBP) reduction. IBP identities relate integrals with different propagator powers and numerators~\cite{Tkachov1981,ChetyrkinTkachov1981,Laporta2000}, and the number of independent master integrals in a fixed family is finite~\cite{SmirnovPetukhov2011}. The coefficients studied here belong to the preceding partial-fraction step and are independent of loop momenta.

Within this framework, the determinantal coefficient formula proved here, reproduced in Eq.~\eqref{eq:2.8}, is a central algebraic ingredient of the explicit partial-fraction reduction. It determines the complete coefficient associated with each reduced denominator set, thereby establishing an explicit connection between the original integrand and the finite basis topologies. A rigorous proof of this formula therefore provides an algebraic foundation for this key step in high-multiplicity reduction and clarifies the conditions under which the representation is valid.

One way to construct this decomposition is to solve an affine relation with a nonzero constant term for \(1\). Dividing by the product of denominators then cancels one denominator in each term, and the operation can be repeated on the remaining relations. More general multivariate partial-fraction algorithms use polynomial combinations of denominators and account for both common zeros and algebraic dependencies~\cite{Leinartas1978,Raichev2012,HellerManteuffel2022}. For linear denominators, existence criteria can also be formulated in terms of their hyperplane arrangement~\cite{deKorteYu2026}. Partial fractions are also used to simplify the rational functions of kinematic invariants that occur as IBP reduction coefficients~\cite{Pak2012,BoehmEtAl2020,HellerManteuffel2022}; in that application, the variables are kinematic invariants rather than loop-dependent scalar products.

A second way to determine coefficients is to evaluate the terms of the decomposition by setting selected propagators to zero. Quadruple cuts have been successfully applied to obtain the one-loop box coefficients in~\cite{BrittoCachazoFeng2005}. Then different multiple cuts have been systematically established to evaluate various coefficients of master integrals for one-loop amplitudes in~\cite{OPP2007}. Numerical-unitarity and Laurent-expansion methods likewise use cut data to determine one-loop coefficients~\cite{OPP2007,EllisGieleKunszt2008,MastroliaMirabellaPeraro2012}. The generalized unitarity method is further developed in~\cite{Forde2007,GieleKunsztMelnikov2008}.

At two loops, cut residues are parameterized by polynomial numerator bases in irreducible scalar products~\cite{MastroliaOssola2011,BadgerFrellesvigZhang2012,FengHuang2013}; the hepta-cut construction and basis classification in Refs.~\cite{BadgerFrellesvigZhang2012,FengHuang2013} use strictly four-dimensional loop momenta. Multivariate polynomial division and computational algebraic geometry extend the decomposition to more general multiloop integrands~\cite{Mastrolia2012,Zhang2012,Mastrolia2013}, including repeated propagators in the dimensionally regulated recursion of Ref.~\cite{Mastrolia2013}. For affine-linear denominators, LinApart3~\cite{FekeshazyKardos2026} extracts partial-fraction coefficients from residues in denominator coordinates. With unit propagator powers, this residue extraction reduces to evaluating the nonsingular factor at the common zero of the retained denominators.

In this work, we give two independent proofs of the determinantal coefficient formula in Eq.~\eqref{eq:2.8}, introduced by Bargie\l a and Yang~\cite{BargielaYang2024}.
First we review the known tHV reduction coefficients for unit propagator powers from the affine relations specified in Sec.~\ref{sec:setup}. In Sec.~\ref{sec:recursive}, we provide an induction proof by summing the ordered denominator-removal paths.
In Sec.~\ref{sec:cuts} we give a second proof by cut evaluation: after clearing the common denominator, a complementary cut leaves one nonzero product, whose reciprocal is the required coefficient. The number of denominators set to zero follows from the affine relations. In Sec.~\ref{sec:degenerate} we examine vanishing determinants and give an example in which terms with divergent coefficients combine into a finite polynomial identity as the product basis degenerates. These limits are taken before integration.

\section{Algebraic setup and reduction formula}\label{sec:setup}

Let us consider a scalar \(L\)-loop integrand
\begin{equation}
\mathcal I(\ell_1,\ldots,\ell_L)
=
\frac{\mathcal N(\ell_1,\ldots,\ell_L)}
{\prod_{j=1}^{m}D_j},
\label{eq:2.1}
\end{equation}
where the \(D_j\) are standard quadratic propagator denominators, each appearing to the first power in Eq.~\eqref{eq:2.1}. Suppose that the external momenta span a \(d_0\)-dimensional space, with basis \(p_1,\ldots,p_{d_0}\). All other external momenta are linear combinations of these vectors, so every scalar product between a loop momentum and an external momentum is a linear combination of \(\ell_a\cdot p_i\). When the propagator denominators are expanded, the loop-momentum-dependent terms are linear combinations of \(\ell_a\cdot\ell_b\) and \(\ell_a\cdot p_i\), with \(1\leq a\leq b\leq L\) and \(1\leq i\leq d_0\); the constant terms depend only on external kinematics and masses. There are \(L(L+1)/2\) and \(Ld_0\) scalar products of the two types, respectively. Each denominator is therefore an affine-linear function of this finite set of scalar products~\cite{BargielaYang2024}. For \(L=2\) and \(d_0=4\), the count gives eleven scalar products, as already used in the two-loop reduction of Ref.~\cite{GluzaKajdaKosower2011}.

For an \(L\)-loop, \(n\)-point topology with trivalent vertices, the maximal number $\mathscr D(n,L)$ of denominators grows with \(n\). Denoting by \(N(L,d_0)\) the multiplicity at which this number equals the scalar-product count gives~\cite{BargielaYang2024}
\begin{equation}
\mathscr D(n,L)=n+3(L-1),
\qquad
N(L,d_0)
=
\frac{L^2+2Ld_0-5L+6}{2}.
\label{eq:2.2}
\end{equation}
The scalar-product count thus bounds the number of independent denominators by
\begin{equation}
r=\mathscr D\bigl(N(L,d_0),L\bigr)=\frac{L(L+1)}{2}+Ld_0,
\qquad
A=m-r.
\label{eq:2.3}
\end{equation}

We first consider denominator families that attain this bound: \(r\) denominators can be chosen so that they are linearly independent together with the constant \(1\), and all remaining denominators are affine combinations of them. When \(m>r\), there are \(A=m-r\) independent affine relations. The reduction removes \(A\) denominators from each term, leaving \(r\). If the actual number of independent denominators is smaller than the bound, \(r\), \(A\), and the relation system are instead defined using that actual number.

Linear elimination of the scalar-product coefficients produces \(A\) independent combinations in which all loop-momentum-dependent terms cancel. The remaining terms depend only on external kinematics and masses. Moving these constant terms to the same side, we write a basis of the complete relation space as~\cite{BargielaYang2024}
\begin{equation}
\alpha_{\ell0}
+\sum_{j=1}^{m}\alpha_{\ell j}D_j
=0,
\qquad
\ell=1,\ldots,A.
\label{eq:2.4}
\end{equation}
The coefficients \(\alpha_{\ell j}\) are rational functions of external kinematic invariants and masses and are independent of loop momenta. All algebraic operations below are performed over the rational function field generated by these kinematic variables. The denominator coefficient matrix has \(A\) rows and \(m\) columns, with \(\ell=1,\ldots,A\) and \(j=1,\ldots,m\). Its rank is \(A\), and Eq.~\eqref{eq:2.4} defines an affine space \(\mathcal V\) of dimension \(r=m-A\) in the coordinates \(D_1,\ldots,D_m\).

For any ordered index tuple \((j_1,\ldots,j_A)\) drawn from \(\{0,1,\ldots,m\}\), we define
\begin{equation}
B_{j_1,\ldots,j_A}
=
\det
\begin{pmatrix}
\alpha_{1j_1}&\cdots&\alpha_{1j_A}\\
\vdots&&\vdots\\
\alpha_{Aj_1}&\cdots&\alpha_{Aj_A}
\end{pmatrix}.
\label{eq:2.5}
\end{equation}
Here, the index \(0\) denotes the constant column.
The proofs in Secs.~\ref{sec:recursive} and~\ref{sec:cuts} use the following nondegeneracy assumption: for every \(1\leq i_1<\cdots<i_A\leq m\),
\begin{equation}
B_{i_1,\ldots,i_A}\neq0,
\qquad
B_{i_1,\ldots,i_{t-1},0,i_{t+1},\ldots,i_A}\neq0
\quad(t=1,\ldots,A).
\label{eq:2.6}
\end{equation}
In Eq.~\eqref{eq:2.6}, the determinant without the index \(0\) ensures that the complementary cut equations have a unique solution, while the determinants with the index \(0\) ensure that the remaining \(A\) denominators do not vanish there. This is an additional assumption on the denominator family. If some of these determinants vanish identically because of the structure of the denominator family, varying the kinematic parameters alone cannot satisfy it. Section~\ref{sec:degenerate} discusses the behavior of the coefficient formula when determinants approach zero.

The partial-fraction decomposition to be established is
\begin{equation}
\frac{1}{\prod_{j=1}^{m}D_j}
=
\sum_{1\leq i_1<\cdots<i_A\leq m}
\frac{c_{i_1,\ldots,i_A}}
{\displaystyle
\prod_{\substack{j=1\\j\notin\{i_1,\ldots,i_A\}}}^{m}D_j},
\label{eq:2.7}
\end{equation}
where \(i_1,\ldots,i_A\) label the denominators removed in the corresponding term. Each term removes exactly \(A\) denominators and retains \(m-A=r\) propagators. The coefficient formula given in Ref.~\cite{BargielaYang2024}, Eq.~(15), is
\begin{equation}
\boxed{
c_{i_1,\ldots,i_A}
=
\frac{\bigl(-B_{i_1,\ldots,i_A}\bigr)^A}
{\displaystyle
\prod_{t=1}^{A}
B_{i_1,\ldots,i_{t-1},0,i_{t+1},\ldots,i_A}}.
}
\label{eq:2.8}
\end{equation}
Multiplying the proposed decomposition~\eqref{eq:2.7} by the full common denominator reduces the problem to establishing the polynomial identity on \(\mathcal V\):
\begin{equation}
\boxed{
1
=
\sum_{1\leq i_1<\cdots<i_A\leq m}
c_{i_1,\ldots,i_A}D_{i_1}\cdots D_{i_A}.
}
\label{eq:2.9}
\end{equation}
Each product \(c_{i_1,\ldots,i_A}D_{i_1}\cdots D_{i_A}\) in Eq.~\eqref{eq:2.9} is the corresponding term of Eq.~\eqref{eq:2.7} obtained by multiplying all denominators. The \(A\) factors in that product are precisely the denominators removed from the term. This correspondence motivates the complementary cuts used in Sec.~\ref{sec:cuts}.

Equations~\eqref{eq:2.7} and~\eqref{eq:2.9} are equivalent on the dense open set where all denominators are nonzero. The aim of this paper is to give a rigorous proof of them. Once Eq.~\eqref{eq:2.9} holds on all of \(\mathcal V\), dividing by the common denominator gives the desired rational-function identity. A general numerator \(\mathcal N(\ell_1,\ldots,\ell_L)\) can be included by multiplying both sides of Eq.~\eqref{eq:2.7}. If it contains generalized propagators expressible in terms of the denominators, the corresponding tHV numerator reduction can be performed before or after the partial-fraction step~\cite{BargielaYang2024}.

\section{Proof by recursive elimination}\label{sec:recursive}
We first prove Eq.~\eqref{eq:2.8} by applying successive partial-fraction decompositions to the affine relations~\eqref{eq:2.4}. The calculation is initially performed where all constant terms and elimination pivots used in the recursion are nonzero. The choice of relation basis and the extension of the resulting formula are discussed at the end of this section.

Take the first relation in Eq.~\eqref{eq:2.4}:
\begin{equation}
\alpha_{10}+\sum_{p=1}^{m}\alpha_{1p}D_p=0.
\label{eq:3.1}
\end{equation}
If \(\alpha_{10}\neq0\), solving Eq.~\eqref{eq:3.1} for the constant \(1\) gives
\begin{equation}
1=-\frac{1}{\alpha_{10}}\sum_{p=1}^{m}\alpha_{1p}D_p.
\label{eq:3.2}
\end{equation}
Multiplying both sides of Eq.~\eqref{eq:3.2} by \(1/\prod_{p=1}^{m}D_p\) and canceling \(D_p\) in each term yields the first partial-fraction step,
\begin{equation}
\frac1{\prod_{p=1}^{m}D_p}
=
\sum_{p=1}^{m}
\left(-\frac{\alpha_{1p}}{\alpha_{10}}\right)
\frac1{\displaystyle
\prod_{\substack{q=1\\q\neq p}}^{m}D_q}.
\label{eq:3.3}
\end{equation}
Equations~\eqref{eq:3.2} and~\eqref{eq:3.3} implement the standard partial-fraction step of expressing \(1\) as a combination of denominators and then canceling one factor in each term~\cite{Leinartas1978,Raichev2012,HellerManteuffel2022}. For loop integrands, related constructions resolve dependent propagators and reduce denominator sets~\cite{Kleiss2012,Pak2012,BargielaYang2024}. Here the affine relations provide coefficients independent of loop momenta, so this step requires only linear algebra.

The \(p\)th term in Eq.~\eqref{eq:3.3} removes \(D_p\) with coefficient \(-\alpha_{1p}/\alpha_{10}\). After each removal, its coefficient in the relation is used as a pivot to eliminate that denominator from the remaining affine relations. The same operation is then repeated on the smaller system.

For a fixed set of \(A\) denominators to be removed, the possible removal orders form \(A!\) ordered paths. Each path contributes the product of its successive coefficients, and the complete reduction coefficient is the sum of all path contributions. We show that this sum is Eq.~\eqref{eq:2.8}. To get the idea, first we display how the six paths combine for \(A=3\), and then prove the general case by induction.

\Needspace{20\baselineskip}
\subsection{The case of three relations}\label{sec:three}

For \(A=3\), the three independent affine relations are
\begin{subequations}
\begin{equation}
\alpha_{10}+\sum_{p=1}^{m}\alpha_{1p}D_p=0.
\label{eq:3.4a}
\end{equation}
\begin{equation}
\alpha_{20}+\sum_{p=1}^{m}\alpha_{2p}D_p=0.
\label{eq:3.4b}
\end{equation}
\begin{equation}
\alpha_{30}+\sum_{p=1}^{m}\alpha_{3p}D_p=0.
\label{eq:3.4c}
\end{equation}
\end{subequations}

Fix the denominators \(D_i,D_j,D_k\) to be removed, with \(i<j<k\), and first consider the ordered path \(i\to j\to k\). Separating the \(p=i\) term in Eq.~\eqref{eq:3.3} gives

\begin{equation}
\frac1{\displaystyle\prod_{p=1}^{m}D_p}
=
-\frac{\alpha_{1i}}{\alpha_{10}}
\frac1{\displaystyle
\prod_{\substack{p=1\\p\neq i}}^{m}D_p}
+
\sum_{\substack{q=1\\q\neq i}}^{m}
\left(-\frac{\alpha_{1q}}{\alpha_{10}}\right)
\frac1{\displaystyle
\prod_{\substack{p=1\\p\neq q}}^{m}D_p}.
\label{eq:3.5}
\end{equation}

The first term on the right of Eq.~\eqref{eq:3.5} removes \(D_i\), contributing \(-\alpha_{1i}/\alpha_{10}\). The other terms enter the final sum over the first removal. Subtracting \(\alpha_{2i}/\alpha_{1i}\) and \(\alpha_{3i}/\alpha_{1i}\) times Eq.~\eqref{eq:3.4a} from Eqs.~\eqref{eq:3.4b} and~\eqref{eq:3.4c}, respectively, eliminates \(D_i\) and gives
\begin{subequations}
\begin{equation}
\left(
\alpha_{20}-\frac{\alpha_{2i}}{\alpha_{1i}}\alpha_{10}
\right)
+
\left(
\alpha_{2j}-\frac{\alpha_{2i}}{\alpha_{1i}}\alpha_{1j}
\right)D_j
+\sum_{\substack{p=1\\p\notin\{i,j\}}}^{m}
\left(
\alpha_{2p}-\frac{\alpha_{2i}}{\alpha_{1i}}\alpha_{1p}
\right)D_p=0.
\label{eq:3.6a}
\end{equation}
\begin{equation}
\left(
\alpha_{30}-\frac{\alpha_{3i}}{\alpha_{1i}}\alpha_{10}
\right)
+
\left(
\alpha_{3j}-\frac{\alpha_{3i}}{\alpha_{1i}}\alpha_{1j}
\right)D_j
+\sum_{\substack{p=1\\p\notin\{i,j\}}}^{m}
\left(
\alpha_{3p}-\frac{\alpha_{3i}}{\alpha_{1i}}\alpha_{1p}
\right)D_p=0.
\label{eq:3.6b}
\end{equation}
\end{subequations}
Solving Eq.~\eqref{eq:3.6a} for \(1\), multiplying by \(1/\prod_{\substack{p=1\\p\neq i}}^{m}D_p\), and separating the \(q=j\) term yields
\begin{equation}
\begin{aligned}
\frac{1}{\displaystyle
\prod_{\substack{p=1\\p\neq i}}^{m}D_p}
&=
-\frac{
\alpha_{1i}\alpha_{2j}-\alpha_{2i}\alpha_{1j}}
{\alpha_{1i}\alpha_{20}-\alpha_{2i}\alpha_{10}}
\frac{1}{\displaystyle
\prod_{\substack{p=1\\p\notin\{i,j\}}}^{m}D_p}
\quad+
\sum_{\substack{q=1\\q\notin\{i,j\}}}^{m}
\left[
-\frac{
\alpha_{1i}\alpha_{2q}-\alpha_{2i}\alpha_{1q}}
{\alpha_{1i}\alpha_{20}-\alpha_{2i}\alpha_{10}}
\right]
\frac{1}{\displaystyle
\prod_{\substack{p=1\\p\notin\{i,q\}}}^{m}D_p}.
\end{aligned}
\label{eq:3.7}
\end{equation}
The first term on the right of Eq.~\eqref{eq:3.7} corresponds to \(i\to j\); its coefficient is the second-step factor. To eliminate \(D_j\), subtract Eq.~\eqref{eq:3.6a} multiplied by the ratio of the \(D_j\) coefficients in Eqs.~\eqref{eq:3.6b} and~\eqref{eq:3.6a} from Eq.~\eqref{eq:3.6b}. Explicitly, the resulting relation is
\begin{equation}
\begin{aligned}
&\left(
\alpha_{30}-\frac{\alpha_{3i}}{\alpha_{1i}}\alpha_{10}
\right)
-\frac{
\alpha_{3j}-\dfrac{\alpha_{3i}}{\alpha_{1i}}\alpha_{1j}}
{\alpha_{2j}-\dfrac{\alpha_{2i}}{\alpha_{1i}}\alpha_{1j}}
\left(
\alpha_{20}-\frac{\alpha_{2i}}{\alpha_{1i}}\alpha_{10}
\right)
\\
&\quad+
\left[
\left(
\alpha_{3k}-\frac{\alpha_{3i}}{\alpha_{1i}}\alpha_{1k}
\right)
-\frac{
\alpha_{3j}-\dfrac{\alpha_{3i}}{\alpha_{1i}}\alpha_{1j}}
{\alpha_{2j}-\dfrac{\alpha_{2i}}{\alpha_{1i}}\alpha_{1j}}
\left(
\alpha_{2k}-\frac{\alpha_{2i}}{\alpha_{1i}}\alpha_{1k}
\right)
\right]D_k
\\
&\quad+
\sum_{\substack{p=1\\p\notin\{i,j,k\}}}^{m}
\left[
\left(
\alpha_{3p}-\frac{\alpha_{3i}}{\alpha_{1i}}\alpha_{1p}
\right)
-\frac{
\alpha_{3j}-\dfrac{\alpha_{3i}}{\alpha_{1i}}\alpha_{1j}}
{\alpha_{2j}-\dfrac{\alpha_{2i}}{\alpha_{1i}}\alpha_{1j}}
\left(
\alpha_{2p}-\frac{\alpha_{2i}}{\alpha_{1i}}\alpha_{1p}
\right)
\right]D_p
=0.
\end{aligned}
\label{eq:3.8}
\end{equation}
Solving Eq.~\eqref{eq:3.8} for \(1\), multiplying by \(1/\prod_{\substack{p=1\\p\notin\{i,j\}}}^{m}D_p\), and separating the \(q=k\) term gives
\begin{equation}
\begin{aligned}
\frac{1}{\displaystyle
\prod_{\substack{p=1\\p\notin\{i,j\}}}^{m}D_p}
&=
-\frac{
\left(\alpha_{3k}-\dfrac{\alpha_{3i}}{\alpha_{1i}}\alpha_{1k}\right)
-\dfrac{
\alpha_{3j}-\dfrac{\alpha_{3i}}{\alpha_{1i}}\alpha_{1j}}
{\alpha_{2j}-\dfrac{\alpha_{2i}}{\alpha_{1i}}\alpha_{1j}}
\left(\alpha_{2k}-\dfrac{\alpha_{2i}}{\alpha_{1i}}\alpha_{1k}\right)
}{
\left(\alpha_{30}-\dfrac{\alpha_{3i}}{\alpha_{1i}}\alpha_{10}\right)
-\dfrac{
\alpha_{3j}-\dfrac{\alpha_{3i}}{\alpha_{1i}}\alpha_{1j}}
{\alpha_{2j}-\dfrac{\alpha_{2i}}{\alpha_{1i}}\alpha_{1j}}
\left(\alpha_{20}-\dfrac{\alpha_{2i}}{\alpha_{1i}}\alpha_{10}\right)
}
\frac{1}{\displaystyle
\prod_{\substack{p=1\\p\notin\{i,j,k\}}}^{m}D_p}
\\
&\quad+
\sum_{\substack{q=1\\q\notin\{i,j,k\}}}^{m}
\left[
-\frac{
\left(\alpha_{3q}-\dfrac{\alpha_{3i}}{\alpha_{1i}}\alpha_{1q}\right)
-\dfrac{
\alpha_{3j}-\dfrac{\alpha_{3i}}{\alpha_{1i}}\alpha_{1j}}
{\alpha_{2j}-\dfrac{\alpha_{2i}}{\alpha_{1i}}\alpha_{1j}}
\left(\alpha_{2q}-\dfrac{\alpha_{2i}}{\alpha_{1i}}\alpha_{1q}\right)
}{
\left(\alpha_{30}-\dfrac{\alpha_{3i}}{\alpha_{1i}}\alpha_{10}\right)
-\dfrac{
\alpha_{3j}-\dfrac{\alpha_{3i}}{\alpha_{1i}}\alpha_{1j}}
{\alpha_{2j}-\dfrac{\alpha_{2i}}{\alpha_{1i}}\alpha_{1j}}
\left(\alpha_{20}-\dfrac{\alpha_{2i}}{\alpha_{1i}}\alpha_{10}\right)
}
\right]
\frac{1}{\displaystyle
\prod_{\substack{p=1\\p\notin\{i,j,q\}}}^{m}D_p}.
\end{aligned}
\label{eq:3.9}
\end{equation}
The first term on the right of Eq.~\eqref{eq:3.9} corresponds to \(i\to j\to k\). Its third-step factor is
\begin{equation}
-\frac{
\left(\alpha_{3k}-\dfrac{\alpha_{3i}}{\alpha_{1i}}\alpha_{1k}\right)
-\dfrac{
\alpha_{3j}-\dfrac{\alpha_{3i}}{\alpha_{1i}}\alpha_{1j}}
{\alpha_{2j}-\dfrac{\alpha_{2i}}{\alpha_{1i}}\alpha_{1j}}
\left(\alpha_{2k}-\dfrac{\alpha_{2i}}{\alpha_{1i}}\alpha_{1k}\right)
}{
\left(\alpha_{30}-\dfrac{\alpha_{3i}}{\alpha_{1i}}\alpha_{10}\right)
-\dfrac{
\alpha_{3j}-\dfrac{\alpha_{3i}}{\alpha_{1i}}\alpha_{1j}}
{\alpha_{2j}-\dfrac{\alpha_{2i}}{\alpha_{1i}}\alpha_{1j}}
\left(\alpha_{20}-\dfrac{\alpha_{2i}}{\alpha_{1i}}\alpha_{10}\right)
}
=
-\frac{B_{i,j,k}}{B_{i,j,0}}.
\label{eq:3.10}
\end{equation}
In Eq.~\eqref{eq:3.8}, the coefficient of \(D_k\) and the constant term share the denominator \(\alpha_{1i}\alpha_{2j}-\alpha_{2i}\alpha_{1j}\). Their numerators are \(B_{i,j,k}\) and \(B_{i,j,0}\), respectively. Solving for \(1\) gives the negative ratio of these quantities, which is Eq.~\eqref{eq:3.10}.

The three factors for \(i\to j\to k\) are given by Eqs.~\eqref{eq:3.5}, \eqref{eq:3.7}, and~\eqref{eq:3.10}. Their product contributes
\begin{equation}
-\frac{\alpha_{1i}}{\alpha_{10}}
\frac{
\alpha_{1i}\alpha_{2j}-\alpha_{2i}\alpha_{1j}}
{\alpha_{1i}\alpha_{20}-\alpha_{2i}\alpha_{10}}
\frac{B_{i,j,k}}{B_{i,j,0}}.
\label{eq:3.11}
\end{equation}
to \(c_{i,j,k}\). The other path beginning with \(D_i\) is \(i\to k\to j\). Simultaneously interchanging \(j\) and \(k\) in Eq.~\eqref{eq:3.11} gives its contribution,
\begin{equation}
-\frac{\alpha_{1i}}{\alpha_{10}}
\frac{
\alpha_{1i}\alpha_{2k}-\alpha_{2i}\alpha_{1k}}
{\alpha_{1i}\alpha_{20}-\alpha_{2i}\alpha_{10}}
\frac{B_{i,k,j}}{B_{i,k,0}}.
\label{eq:3.12}
\end{equation}
With \(B_{i,k,j}=-B_{i,j,k}\) and \(B_{i,k,0}=-B_{i,0,k}\), the sum of Eqs.~\eqref{eq:3.11} and~\eqref{eq:3.12} is simplified by the identity
\begin{equation}
\begin{aligned}
&\left(\alpha_{1i}\alpha_{2j}-\alpha_{2i}\alpha_{1j}\right)B_{i,0,k}
+\left(\alpha_{1i}\alpha_{2k}-\alpha_{2i}\alpha_{1k}\right)B_{i,j,0}
\\
&\hspace{35mm}=
\left(\alpha_{1i}\alpha_{20}-\alpha_{2i}\alpha_{10}\right)B_{i,j,k}
\end{aligned}
\label{eq:3.13}
\end{equation}
Equation~\eqref{eq:3.13} follows by expanding the three determinants. Substituting it into the sum of Eqs.~\eqref{eq:3.11} and~\eqref{eq:3.12} cancels the common factor \(\alpha_{1i}\alpha_{20}-\alpha_{2i}\alpha_{10}\), giving
\begin{equation}
-\frac{\alpha_{1i}}{\alpha_{10}}
\frac{B_{i,j,k}^{\,2}}
{B_{i,0,k}B_{i,j,0}}.
\label{eq:3.14}
\end{equation}
If \(D_j\) is removed first, subtract \(\alpha_{2j}/\alpha_{1j}\) and \(\alpha_{3j}/\alpha_{1j}\) times Eq.~\eqref{eq:3.4a} from Eqs.~\eqref{eq:3.4b} and~\eqref{eq:3.4c}, respectively. The paths \(j\to i\to k\) and \(j\to k\to i\) then sum to
\begin{equation}
-\frac{\alpha_{1j}}{\alpha_{10}}
\frac{B_{i,j,k}^{\,2}}
{B_{0,j,k}B_{i,j,0}}.
\label{eq:3.15}
\end{equation}
Likewise, removing \(D_k\) first requires subtracting \(\alpha_{2k}/\alpha_{1k}\) and \(\alpha_{3k}/\alpha_{1k}\) times Eq.~\eqref{eq:3.4a} from Eqs.~\eqref{eq:3.4b} and~\eqref{eq:3.4c}. The paths \(k\to i\to j\) and \(k\to j\to i\) sum to
\begin{equation}
-\frac{\alpha_{1k}}{\alpha_{10}}
\frac{B_{i,j,k}^{\,2}}
{B_{0,j,k}B_{i,0,k}}.
\label{eq:3.16}
\end{equation}
Adding Eqs.~\eqref{eq:3.14}--\eqref{eq:3.16} and writing all determinants in the common column order \(i<j<k\) gives
\begin{equation}
\begin{aligned}
c_{i,j,k}
&=
-\frac{B_{i,j,k}^{\,2}}
{\alpha_{10}B_{0,j,k}B_{i,0,k}B_{i,j,0}}
\\
&\quad\times
\left(
\alpha_{1i}B_{0,j,k}
+\alpha_{1j}B_{i,0,k}
+\alpha_{1k}B_{i,j,0}
\right).
\end{aligned}
\label{eq:3.17}
\end{equation}
The three terms in parentheses arise from the pairs of paths beginning with \(D_i,D_j,D_k\), respectively, and satisfy
\begin{equation}
\alpha_{1i}B_{0,j,k}
+\alpha_{1j}B_{i,0,k}
+\alpha_{1k}B_{i,j,0}
=
\alpha_{10}B_{i,j,k}.
\label{eq:3.18}
\end{equation}

To obtain Eq.~\eqref{eq:3.18}, expand each replacement determinant along its constant column. The cross terms containing \(\alpha_{20}\) and \(\alpha_{30}\) cancel pairwise, leaving \(\alpha_{10}B_{i,j,k}\). Substituting into Eq.~\eqref{eq:3.17} yields

\begin{equation}
\boxed{
c_{i,j,k}
=
\frac{(-B_{i,j,k})^3}
{B_{0,j,k}B_{i,0,k}B_{i,j,0}}.
}
\label{eq:3.19}
\end{equation}
The six paths split into three pairs according to the first removed denominator. Within each pair, the last two steps combine in the smaller system; the cofactor identity then sums the three possible first removals. This grouping gives the recurrence used in the general proof.

\Needspace{15\baselineskip}
\subsection{The general case}\label{sec:general}

For general \(A\), write the independent affine relations as
\begin{equation}
\begin{aligned}
\alpha_{10}+\sum_{p=1}^{m}\alpha_{1p}D_p&=0,\\
\alpha_{20}+\sum_{p=1}^{m}\alpha_{2p}D_p&=0,\\
&\ \vdots\\
\alpha_{A0}+\sum_{p=1}^{m}\alpha_{Ap}D_p&=0.
\end{aligned}
\label{eq:3.20}
\end{equation}
Equation~\eqref{eq:3.3} supplies the base case \(A=1\), whose coefficients agree with Eq.~\eqref{eq:2.8}. Assume that Eq.~\eqref{eq:2.8} holds for systems of the same type with \(A-1\) relations spanning their complete affine relation space. Fix the denominators \(D_{i_1},\ldots,D_{i_A}\) to be removed, with \(i_1<\cdots<i_A\).

The \(A!\) ordered paths fall into \(A\) groups according to the first denominator removed. Let \(\widetilde c^{(t)}_{i_1,\ldots,i_{t-1},i_{t+1},\ldots,i_A}\) denote the coefficient for removing the remaining \(A-1\) target denominators in the smaller system after \(D_{i_t}\) has been removed. The first-step factor is given by Eq.~\eqref{eq:3.3}, so the desired coefficient satisfies the recurrence
\begin{equation}
c_{i_1,\ldots,i_A}
=
\sum_{t=1}^{A}
\left(-\frac{\alpha_{1i_t}}{\alpha_{10}}\right)
\widetilde c^{(t)}_{i_1,\ldots,i_{t-1},i_{t+1},\ldots,i_A}.
\label{eq:3.21}
\end{equation}
Fix \(t\). We eliminate \(D_{i_t}\), apply the induction hypothesis to the remaining \(A-1\) relations, and express the resulting determinants in terms of the original \(A\)-row system before substituting into Eq.~\eqref{eq:3.21}. For \(\ell=2,\ldots,A\), subtract \(\alpha_{\ell i_t}/\alpha_{1i_t}\) times the first relation in Eq.~\eqref{eq:3.20} from the \(\ell\)th relation. This gives
\begin{equation}
\left(
\alpha_{\ell0}
-\frac{\alpha_{\ell i_t}}{\alpha_{1i_t}}\alpha_{10}
\right)
+\sum_{\substack{p=1\\p\neq i_t}}^{m}
\left(
\alpha_{\ell p}
-\frac{\alpha_{\ell i_t}}{\alpha_{1i_t}}\alpha_{1p}
\right)D_p
=0,
\qquad \ell=2,\ldots,A.
\label{eq:3.22}
\end{equation}
The nonzero pivot allows \(D_{i_t}\) to be solved for in the first relation. The remaining relations~\eqref{eq:3.22} therefore span the complete affine relation space among the other \(m-1\) denominators. There are \(A-1\) independent relations, with \((m-1)-(A-1)=r\). Applying the induction hypothesis gives
\begin{equation}
c_{j_1,\ldots,j_{A-1}}
=
\frac{\bigl(-B_{j_1,\ldots,j_{A-1}}\bigr)^{A-1}}
{\displaystyle
\prod_{s=1}^{A-1}
B_{j_1,\ldots,j_{s-1},0,j_{s+1},\ldots,j_{A-1}}}.
\label{eq:3.23}
\end{equation}
Here both \(c_{j_1,\ldots,j_{A-1}}\) and \(B\) refer to the smaller system~\eqref{eq:3.22}. Choose the ordered indices \(j_1,\ldots,j_{A-1}\) to be \(i_1,\ldots,i_{t-1},i_{t+1},\ldots,i_A\). The left-hand side of Eq.~\eqref{eq:3.23} is then precisely \(\widetilde c^{(t)}_{i_1,\ldots,i_{t-1},i_{t+1},\ldots,i_A}\) in Eq.~\eqref{eq:3.21}; it includes all \((A-1)!\) paths after the first removal. We now express its unreplaced and replacement determinants in terms of the original \(A\times A\) determinants.

In \(B_{i_1,\ldots,i_A}\), subtract \(\alpha_{\ell i_t}/\alpha_{1i_t}\) times the first row from row \(\ell\), for \(\ell=2,\ldots,A\). The determinant becomes
\begingroup
\fontsize{9}{11}\selectfont
\setlength{\arraycolsep}{2pt}
\renewcommand{\arraystretch}{1.3}
\begin{equation}
\begin{aligned}
B_{i_1,\ldots,i_A}
=\det\!\begin{pmatrix}
\alpha_{1i_1}&\cdots&\alpha_{1i_{t-1}}&\alpha_{1i_t}&\alpha_{1i_{t+1}}&\cdots&\alpha_{1i_A}\\
\alpha_{2i_1}-\tfrac{\alpha_{2i_t}}{\alpha_{1i_t}}\alpha_{1i_1}
&\cdots
&\alpha_{2i_{t-1}}-\tfrac{\alpha_{2i_t}}{\alpha_{1i_t}}\alpha_{1i_{t-1}}
&0
&\alpha_{2i_{t+1}}-\tfrac{\alpha_{2i_t}}{\alpha_{1i_t}}\alpha_{1i_{t+1}}
&\cdots
&\alpha_{2i_A}-\tfrac{\alpha_{2i_t}}{\alpha_{1i_t}}\alpha_{1i_A}\\
\vdots&&\vdots&\vdots&\vdots&&\vdots\\
\alpha_{Ai_1}-\tfrac{\alpha_{Ai_t}}{\alpha_{1i_t}}\alpha_{1i_1}
&\cdots
&\alpha_{Ai_{t-1}}-\tfrac{\alpha_{Ai_t}}{\alpha_{1i_t}}\alpha_{1i_{t-1}}
&0
&\alpha_{Ai_{t+1}}-\tfrac{\alpha_{Ai_t}}{\alpha_{1i_t}}\alpha_{1i_{t+1}}
&\cdots
&\alpha_{Ai_A}-\tfrac{\alpha_{Ai_t}}{\alpha_{1i_t}}\alpha_{1i_A}
\end{pmatrix}.
\end{aligned}
\label{eq:3.24}
\end{equation}
\endgroup
Only \(\alpha_{1i_t}\) remains nonzero in column \(t\) of Eq.~\eqref{eq:3.24}. Expanding along this column gives the unreplaced determinant in the numerator of Eq.~\eqref{eq:3.23}:
\begingroup
\fontsize{9}{11}\selectfont
\setlength{\arraycolsep}{2pt}
\renewcommand{\arraystretch}{1.3}
\begin{equation}
\begin{aligned}
&\det\!\begin{pmatrix}
\alpha_{2i_1}-\tfrac{\alpha_{2i_t}}{\alpha_{1i_t}}\alpha_{1i_1}
&\cdots
&\alpha_{2i_{t-1}}-\tfrac{\alpha_{2i_t}}{\alpha_{1i_t}}\alpha_{1i_{t-1}}
&\alpha_{2i_{t+1}}-\tfrac{\alpha_{2i_t}}{\alpha_{1i_t}}\alpha_{1i_{t+1}}
&\cdots
&\alpha_{2i_A}-\tfrac{\alpha_{2i_t}}{\alpha_{1i_t}}\alpha_{1i_A}\\
\vdots&&\vdots&\vdots&&\vdots\\
\alpha_{Ai_1}-\tfrac{\alpha_{Ai_t}}{\alpha_{1i_t}}\alpha_{1i_1}
&\cdots
&\alpha_{Ai_{t-1}}-\tfrac{\alpha_{Ai_t}}{\alpha_{1i_t}}\alpha_{1i_{t-1}}
&\alpha_{Ai_{t+1}}-\tfrac{\alpha_{Ai_t}}{\alpha_{1i_t}}\alpha_{1i_{t+1}}
&\cdots
&\alpha_{Ai_A}-\tfrac{\alpha_{Ai_t}}{\alpha_{1i_t}}\alpha_{1i_A}
\end{pmatrix}
\\[1mm]
&\hspace{18mm}=
(-1)^{1+t}\frac{B_{i_1,\ldots,i_A}}{\alpha_{1i_t}}.
\end{aligned}
\label{eq:3.25}
\end{equation}
\endgroup
Equation~\eqref{eq:3.25} expresses the smaller-system determinant in terms of \(B_{i_1,\ldots,i_A}\). If \(t=1\) or \(t=A\), the corresponding empty column blocks in Eqs.~\eqref{eq:3.24} and~\eqref{eq:3.25} are omitted.

Next fix \(u\neq t\). For display, first take \(u<t\), with nonadjacent \(u,t\) away from the endpoints. In \(B_{i_1,\ldots,i_{u-1},0,i_{u+1},\ldots,i_A}\), subtract \(\alpha_{\ell i_t}/\alpha_{1i_t}\) times the first row from row \(\ell\), for \(\ell=2,\ldots,A\). This gives
\begingroup
\fontsize{9}{11}\selectfont
\setlength{\arraycolsep}{2pt}
\renewcommand{\arraystretch}{1.3}
\begin{equation}
\begin{aligned}
&B_{i_1,\ldots,i_{u-1},0,i_{u+1},\ldots,i_A}
\\
&=\det\!\begin{pmatrix}
\alpha_{1i_1}
&\cdots
&\alpha_{10}
&\cdots
&\alpha_{1i_{t-1}}
&\alpha_{1i_t}
&\alpha_{1i_{t+1}}
&\cdots
&\alpha_{1i_A}\\
\alpha_{2i_1}-\tfrac{\alpha_{2i_t}}{\alpha_{1i_t}}\alpha_{1i_1}
&\cdots
&\alpha_{20}-\tfrac{\alpha_{2i_t}}{\alpha_{1i_t}}\alpha_{10}
&\cdots
&\alpha_{2i_{t-1}}-\tfrac{\alpha_{2i_t}}{\alpha_{1i_t}}\alpha_{1i_{t-1}}
&0
&\alpha_{2i_{t+1}}-\tfrac{\alpha_{2i_t}}{\alpha_{1i_t}}\alpha_{1i_{t+1}}
&\cdots
&\alpha_{2i_A}-\tfrac{\alpha_{2i_t}}{\alpha_{1i_t}}\alpha_{1i_A}\\
\vdots&&\vdots&&\vdots&\vdots&\vdots&&\vdots\\
\alpha_{Ai_1}-\tfrac{\alpha_{Ai_t}}{\alpha_{1i_t}}\alpha_{1i_1}
&\cdots
&\alpha_{A0}-\tfrac{\alpha_{Ai_t}}{\alpha_{1i_t}}\alpha_{10}
&\cdots
&\alpha_{Ai_{t-1}}-\tfrac{\alpha_{Ai_t}}{\alpha_{1i_t}}\alpha_{1i_{t-1}}
&0
&\alpha_{Ai_{t+1}}-\tfrac{\alpha_{Ai_t}}{\alpha_{1i_t}}\alpha_{1i_{t+1}}
&\cdots
&\alpha_{Ai_A}-\tfrac{\alpha_{Ai_t}}{\alpha_{1i_t}}\alpha_{1i_A}
\end{pmatrix}.
\end{aligned}
\label{eq:3.26}
\end{equation}
\endgroup
Equations~\eqref{eq:3.26} and~\eqref{eq:3.24} differ only in column \(u\); column \(t\) still has \(\alpha_{1i_t}\) as its only nonzero entry. Expanding along that column gives the replacement determinant associated with \(D_{i_u}\) in the denominator of Eq.~\eqref{eq:3.23}:
\begingroup
\fontsize{9}{11}\selectfont
\setlength{\arraycolsep}{2pt}
\renewcommand{\arraystretch}{1.3}
\begin{equation}
\begin{aligned}
&\det\!\begin{pmatrix}
\alpha_{2i_1}-\tfrac{\alpha_{2i_t}}{\alpha_{1i_t}}\alpha_{1i_1}
&\cdots
&\alpha_{20}-\tfrac{\alpha_{2i_t}}{\alpha_{1i_t}}\alpha_{10}
&\cdots
&\alpha_{2i_{t-1}}-\tfrac{\alpha_{2i_t}}{\alpha_{1i_t}}\alpha_{1i_{t-1}}
&\alpha_{2i_{t+1}}-\tfrac{\alpha_{2i_t}}{\alpha_{1i_t}}\alpha_{1i_{t+1}}
&\cdots
&\alpha_{2i_A}-\tfrac{\alpha_{2i_t}}{\alpha_{1i_t}}\alpha_{1i_A}\\
\vdots&&\vdots&&\vdots&\vdots&&\vdots\\
\alpha_{Ai_1}-\tfrac{\alpha_{Ai_t}}{\alpha_{1i_t}}\alpha_{1i_1}
&\cdots
&\alpha_{A0}-\tfrac{\alpha_{Ai_t}}{\alpha_{1i_t}}\alpha_{10}
&\cdots
&\alpha_{Ai_{t-1}}-\tfrac{\alpha_{Ai_t}}{\alpha_{1i_t}}\alpha_{1i_{t-1}}
&\alpha_{Ai_{t+1}}-\tfrac{\alpha_{Ai_t}}{\alpha_{1i_t}}\alpha_{1i_{t+1}}
&\cdots
&\alpha_{Ai_A}-\tfrac{\alpha_{Ai_t}}{\alpha_{1i_t}}\alpha_{1i_A}
\end{pmatrix}
\\[1mm]
&\hspace{18mm}=
(-1)^{1+t}
\frac{
B_{i_1,\ldots,i_{u-1},0,i_{u+1},\ldots,i_A}}
{\alpha_{1i_t}}.
\end{aligned}
\label{eq:3.27}
\end{equation}
\endgroup
Adjacent and endpoint cases are obtained by omitting the appropriate empty column blocks. For \(u>t\), the constant column lies to the right of column \(t\), and the same row elimination and column expansion apply. The order of all remaining columns is preserved, so no additional column-permutation sign occurs. Equation~\eqref{eq:3.27} therefore holds for every \(u\neq t\).

Substituting Eqs.~\eqref{eq:3.25} and~\eqref{eq:3.27} into Eq.~\eqref{eq:3.23} gives the smaller-system coefficient in the recurrence~\eqref{eq:3.21},
\begin{equation}
\begin{aligned}
\widetilde c^{(t)}_{i_1,\ldots,i_{t-1},i_{t+1},\ldots,i_A}
&=\frac{
\left[-(-1)^{1+t}B_{i_1,\ldots,i_A}/\alpha_{1i_t}\right]^{A-1}}
{\displaystyle
\prod_{\substack{u=1\\u\neq t}}^{A}
\left[(-1)^{1+t}
B_{i_1,\ldots,i_{u-1},0,i_{u+1},\ldots,i_A}/\alpha_{1i_t}\right]}
\\[1mm]
&=
\frac{
\left[\dfrac{(-1)^{1+t}}{\alpha_{1i_t}}\right]^{A-1}
\bigl(-B_{i_1,\ldots,i_A}\bigr)^{A-1}}
{\displaystyle
\left[\dfrac{(-1)^{1+t}}{\alpha_{1i_t}}\right]^{A-1}
\prod_{\substack{u=1\\u\neq t}}^{A}
B_{i_1,\ldots,i_{u-1},0,i_{u+1},\ldots,i_A}}
\\[1mm]
&=
\frac{\bigl(-B_{i_1,\ldots,i_A}\bigr)^{A-1}}
{\displaystyle
\prod_{\substack{u=1\\u\neq t}}^{A}
B_{i_1,\ldots,i_{u-1},0,i_{u+1},\ldots,i_A}}.
\end{aligned}
\label{eq:3.28}
\end{equation}
The common factor \((-1)^{1+t}/\alpha_{1i_t}\) in Eqs.~\eqref{eq:3.25} and~\eqref{eq:3.27} occurs \(A-1\) times in both the numerator and denominator of Eq.~\eqref{eq:3.28} and cancels. This coefficient sums all \((A-1)!\) paths following the fixed first removal of \(D_{i_t}\).

Substituting Eq.~\eqref{eq:3.28} into Eq.~\eqref{eq:3.21} sums all \(A!\) removal orders. The denominator of the \(t\)th term contains every replacement determinant except the \(t\)th one. Bringing the terms to a common denominator therefore multiplies that numerator by \(B_{i_1,\ldots,i_{t-1},0,i_{t+1},\ldots,i_A}\), yielding

\begin{equation}
\begin{aligned}
c_{i_1,\ldots,i_A}
&=
\sum_{t=1}^{A}
\left(-\frac{\alpha_{1i_t}}{\alpha_{10}}\right)
\frac{\bigl(-B_{i_1,\ldots,i_A}\bigr)^{A-1}}
{\displaystyle
\prod_{\substack{u=1\\u\neq t}}^{A}
B_{i_1,\ldots,i_{u-1},0,i_{u+1},\ldots,i_A}}
\\
&=
-\frac{\bigl(-B_{i_1,\ldots,i_A}\bigr)^{A-1}}
{\displaystyle
\alpha_{10}
\prod_{u=1}^{A}
B_{i_1,\ldots,i_{u-1},0,i_{u+1},\ldots,i_A}}
\\
&\quad\times
\sum_{t=1}^{A}
\alpha_{1i_t}
B_{i_1,\ldots,i_{t-1},0,i_{t+1},\ldots,i_A}.
\end{aligned}
\label{eq:3.29}
\end{equation}

\Needspace{8\baselineskip}
The sum in Eq.~\eqref{eq:3.29} satisfies the following cofactor identity, which is a component of the standard matrix-adjugate identity~\cite{HornJohnson2013}:
\begin{equation}
\boxed{
\sum_{t=1}^{A}
\alpha_{1i_t}
B_{i_1,\ldots,i_{t-1},0,i_{t+1},\ldots,i_A}
=
\alpha_{10}B_{i_1,\ldots,i_A}.
}
\label{eq:3.30}
\end{equation}
Expand the \(t\)th replacement determinant on the left of Eq.~\eqref{eq:3.30} along its constant column and sum over \(t\). The first entry of that column contributes \(\alpha_{10}B_{i_1,\ldots,i_A}\). Each other entry contributes a determinant with two identical rows and hence vanishes. Substitution into Eq.~\eqref{eq:3.29} gives
\begin{equation}
c_{i_1,\ldots,i_A}
=
\frac{\bigl(-B_{i_1,\ldots,i_A}\bigr)^A}
{\displaystyle
\prod_{u=1}^{A}
B_{i_1,\ldots,i_{u-1},0,i_{u+1},\ldots,i_A}},
\label{eq:3.31}
\end{equation}
which completes the induction.

A generic invertible linear recombination of the relations makes the constant terms and intermediate pivots needed in the recursion nonzero. Such a change of relation basis multiplies every determinant in Eq.~\eqref{eq:2.5} by the same nonzero factor. That factor occurs \(A\) times in both the numerator and denominator of Eq.~\eqref{eq:2.8} and cancels, so the final coefficient is independent of the chosen relation basis. Having established the formula where all intermediate pivots are nonzero, we extend it as a rational identity to every other point at which its right-hand side is defined.

\section{Proof by cut evaluation}\label{sec:cuts}

The proof in Sec.~\ref{sec:recursive} explains how successive partial-fraction decomposition produces the closed-form coefficients through elimination and path summation. The polynomial identity obtained after clearing the common denominator also admits a more intuitive proof: choosing a cut point that leaves only the target term determines its coefficient directly from the value of the corresponding denominator product.

Specifically, multiply both sides of the proposed decomposition~\eqref{eq:2.7} by \(\prod_{j=1}^{m}D_j\). The left-hand side becomes \(1\), while the denominator of each term on the right cancels, leaving the product of the \(A\) denominators removed from that term. The reduction problem is thus to establish the polynomial identity~\eqref{eq:2.9}, repeated here as
\begin{equation}
\boxed{
1
=
\sum_{1\leq i_1<\cdots<i_A\leq m}
c_{i_1,\ldots,i_A}D_{i_1}\cdots D_{i_A}.
}
\label{eq:4.1}
\end{equation}
Equation~\eqref{eq:4.1} is the starting point of this proof, where \(c_{i_1,\ldots,i_A}\) are unknown coefficients independent of loop momenta. In the multiple cut method~\cite{BrittoCachazoFeng2005,OPP2007}, selected propagators are set to zero, so that only the terms of interest survive and their coefficients can be extracted directly.

Here we will follow the same logic. If \(k\) distinct denominators are set to zero, with \(0\leq k\leq m-A\), the products not directly eliminated by these conditions must select their \(A\) factors from the remaining \(m-k\) denominators. There are \(\binom{m-k}{A}\) such products. To retain exactly one, take \(m-k=A\), or \(k=m-A=r\). Setting one more denominator to zero would leave fewer than \(A\) nonzero factors, so every \(A\)-factor product would vanish. Thus \(r=m-A\) is the largest number of zero conditions compatible with retaining a nonzero target product. Equation~\eqref{eq:2.3} further ensures that such solutions exist.

Accordingly, fix the target indices \(1\leq i_1<\cdots<i_A\leq m\) and set all \(m-A=r\) complementary denominators to zero. The associated cut point \(P_{i_1,\ldots,i_A}\) on \(\mathcal V\) is specified by
\begin{equation}
D_j(P_{i_1,\ldots,i_A})=0,
\qquad
1\leq j\leq m,\quad j\notin\{i_1,\ldots,i_A\}.
\label{eq:4.2}
\end{equation}
Every \(A\)-factor product other than \(D_{i_1}\cdots D_{i_A}\) contains at least one complementary denominator and therefore vanishes at this point. Only the term indexed by \(\{i_1,\ldots,i_A\}\) remains in Eq.~\eqref{eq:4.1}.
After the cut conditions~\eqref{eq:4.2} are substituted into the affine relations~\eqref{eq:2.4}, the remaining denominator values satisfy
\begin{equation}
\begin{pmatrix}
\alpha_{1i_1}&\cdots&\alpha_{1i_A}\\
\vdots&&\vdots\\
\alpha_{Ai_1}&\cdots&\alpha_{Ai_A}
\end{pmatrix}
\begin{pmatrix}
D_{i_1}(P_{i_1,\ldots,i_A})\\
\vdots\\
D_{i_A}(P_{i_1,\ldots,i_A})
\end{pmatrix}
=
-\begin{pmatrix}
\alpha_{10}\\
\vdots\\
\alpha_{A0}
\end{pmatrix}.
\label{eq:4.3}
\end{equation}
Recall the assumption~\eqref{eq:2.6} in Sec.~\ref{sec:setup}: the unreplaced determinant and all replacement determinants are nonzero. The former gives a unique solution of Eq.~\eqref{eq:4.3}, which, together with the zero complementary coordinates, uniquely specifies a point on \(\mathcal V\). The latter ensures that each remaining denominator value obtained below is nonzero. These cut points serve only as algebraic evaluation points on \(\mathcal V\). Cramer's rule~\cite{HornJohnson2013} gives
\begin{equation}
D_{i_t}(P_{i_1,\ldots,i_A})
=
-\frac{
B_{i_1,\ldots,i_{t-1},0,i_{t+1},\ldots,i_A}}
{B_{i_1,\ldots,i_A}},
\qquad
t=1,\ldots,A.
\label{eq:4.4}
\end{equation}
All \(A\) values in Eq.~\eqref{eq:4.4} are nonzero. Hence, after any \(r=m-A\) denominators are set to zero, no remaining denominator can also vanish: \(r\) is indeed the maximum number of simultaneously vanishing denominators. The target product is
\begin{equation}
\begin{aligned}
D_{i_1}(P_{i_1,\ldots,i_A})\cdots
D_{i_A}(P_{i_1,\ldots,i_A})
&=
(-1)^A
\frac{
\displaystyle\prod_{t=1}^{A}
B_{i_1,\ldots,i_{t-1},0,i_{t+1},\ldots,i_A}}
{B_{i_1,\ldots,i_A}^{\,A}}
\neq0.
\end{aligned}
\label{eq:4.5}
\end{equation}

Evaluating these denominator products at the complementary cut points gives a diagonal matrix with nonzero diagonal entries, so the products are linearly independent. Each has total degree at most \(A\) on the \(r\)-dimensional affine space \(\mathcal V\), and their number equals the dimension of that polynomial space,
\begin{equation}
\binom{r+A}{A}
=
\binom{m}{A}.
\label{eq:4.6}
\end{equation}
The products therefore form a basis of this space, ensuring that the expansion of \(1\) in Eq.~\eqref{eq:4.1} exists and is unique~\cite{ChungYao1977}.

Evaluating Eq.~\eqref{eq:4.1} at \(P_{i_1,\ldots,i_A}\) leaves only the target term:
\begin{equation}
1
=
c_{i_1,\ldots,i_A}
\prod_{t=1}^{A}D_{i_t}(P_{i_1,\ldots,i_A}).
\label{eq:4.7}
\end{equation}
Substituting Eq.~\eqref{eq:4.5} gives the complete coefficient for an arbitrary set of removed denominators,
\begin{equation}
\boxed{
\begin{aligned}
c_{i_1,\ldots,i_A}
&=\frac{1}{D_{i_1}(P_{i_1,\ldots,i_A})\cdots D_{i_A}(P_{i_1,\ldots,i_A})}
=\frac{\bigl(-B_{i_1,\ldots,i_A}\bigr)^A}
{\displaystyle\prod_{t=1}^{A}B_{i_1,\ldots,i_{t-1},0,i_{t+1},\ldots,i_A}}.
\end{aligned}}
\label{eq:4.8}
\end{equation}

This reproduces Eq.~\eqref{eq:2.8} and completes the proof for all reduction coefficients.

\section{Behavior at degenerate kinematics}\label{sec:degenerate}

We now examine Eq.~\eqref{eq:2.8} as the kinematics approach a point where some of its determinants vanish. If the dimension of the complete affine relation space changes, the reduction must be reconstructed. In the following, the number of relations remains \(A=m-r\), and a relation basis is chosen that is regular and has full row rank near the special point.

\subsection{Three cases for the coefficient formula}

Fix \(i_1<\cdots<i_A\). In the first case, all replacement determinants are nonzero:
\begin{equation}
\prod_{t=1}^{A}
B_{i_1,\ldots,i_{t-1},0,i_{t+1},\ldots,i_A}
\neq0.
\label{eq:5.1}
\end{equation}
The right-hand side of Eq.~\eqref{eq:2.8} is then defined. If \(B_{i_1,\ldots,i_A}\neq0\), the coefficient is finite and nonzero; if \(B_{i_1,\ldots,i_A}=0\), its rational continuation gives \(c_{i_1,\ldots,i_A}=0\). Thus a vanishing unreplaced determinant does not by itself cause a divergence.

In the second case, at least one replacement determinant vanishes but \(B_{i_1,\ldots,i_A}\neq0\). As the kinematics approach the special point from nearby points where the coefficient is defined, the denominator of Eq.~\eqref{eq:2.8} tends to zero while its numerator remains nonzero, so \(c_{i_1,\ldots,i_A}\) diverges. If the \(t\)th replacement determinant vanishes, Eq.~\eqref{eq:4.4} gives \(D_{i_t}(P_{i_1,\ldots,i_A})=0\), and the target product is zero. Every other product contains a complementary denominator and vanishes at the same point. Consequently, the right-hand side of Eq.~\eqref{eq:2.9} is zero for any finite coefficients and cannot equal \(1\). The special point therefore admits no expansion of the prescribed form with finite, loop-momentum-independent coefficients.

In the third case, the unreplaced determinant and one or more replacement determinants vanish simultaneously:
\begin{equation}
B_{i_1,\ldots,i_A}=0,
\qquad
\prod_{t=1}^{A}
B_{i_1,\ldots,i_{t-1},0,i_{t+1},\ldots,i_A}=0.
\label{eq:5.2}
\end{equation}
Equation~\eqref{eq:2.8} then has the form \(0/0\), and the determinant values at the special point alone do not determine the coefficient's limit.

For such a limit, choose a fixed analytic path from the nondegenerate region to the special point and parameterize it by \(\lambda\to0\). Assume that the relevant determinants, computed in the regular relation basis, are analytic in \(\lambda\) and do not vanish identically along this path. Let \(\mu\) and \(\nu_t\) be their lowest nonzero powers, with nonzero leading coefficients \(b\) and \(b_t\), respectively. Equation~\eqref{eq:2.8} gives
\begin{equation}
\begin{aligned}
B_{i_1,\ldots,i_A}(\lambda)
&=
b\lambda^{\mu}\bigl[1+\mathcal O(\lambda)\bigr],
\\
B_{i_1,\ldots,i_{t-1},0,i_{t+1},\ldots,i_A}
(\lambda)
&=
b_t\lambda^{\nu_t}\bigl[1+\mathcal O(\lambda)\bigr],
\qquad t=1,\ldots,A,
\\
c_{i_1,\ldots,i_A}(\lambda)
&=
\frac{(-b)^A}{\displaystyle\prod_{t=1}^{A}b_t}
\lambda^{ A\mu-\sum_{t=1}^{A}\nu_t}
\bigl[1+\mathcal O(\lambda)\bigr],
\qquad \lambda\to0.
\end{aligned}
\label{eq:5.3}
\end{equation}
A negative, zero, or positive exponent gives, respectively, a divergent coefficient, a finite nonzero limit, or a vanishing coefficient. This statement applies along the chosen path. If the dimension of the complete relation space changes, the reduction must first be reconstructed.

In particular, suppose all relations are homogeneous, so that \(\alpha_{\ell0}=0\) for every \(\ell=1,\ldots,A\). All replacement determinants then vanish, although the unreplaced determinants need not. For each fixed index set, this is the second or third case above. At the point \(D_1=\cdots=D_m=0\) of \(\mathcal V\), every product on the right of Eq.~\eqref{eq:2.9} vanishes. No finite-coefficient expansion of the required form can therefore exist. The role of common zeros is central to integrand reducibility tests and multivariate partial-fraction decompositions~\cite{Kleiss2012,Leinartas1978,Raichev2012}.

\subsection{An example of a degenerate limit}

Consider three denominators with two independent affine relations:
\begin{equation}
\begin{gathered}
D_1=x,
\qquad
D_2=2x+\lambda,
\qquad
D_3=1+x,
\\[1mm]
\lambda+2D_1-D_2=0,
\qquad
1+D_1-D_3=0.
\end{gathered}
\label{eq:5.4}
\end{equation}
Here \(D_1=x\) is the only independent denominator variable, and \(D_2,D_3\) are affine functions of it. The minor formed by the \(D_2\) and \(D_3\) columns of the relation coefficient matrix equals \(1\) for all \(\lambda\). Thus the two relations remain independent at \(\lambda=0\) and \(\lambda=2\), and \(A=2\) throughout. At these two points, \(D_2=2D_1\) and \(D_2=2D_3\), respectively, so the denominator-product basis degenerates. For \(\lambda\notin\{0,2\}\), Eq.~\eqref{eq:2.8} gives
\begin{equation}
c_{1,2}=\frac{1}{2-\lambda},
\qquad
c_{1,3}=-\frac{4}{\lambda(2-\lambda)},
\qquad
c_{2,3}=\frac{1}{\lambda}.
\label{eq:5.5}
\end{equation}
The corresponding polynomial identity is
\begin{equation}
1
=
\frac{D_1D_2}{2-\lambda}
-\frac{4D_1D_3}{\lambda(2-\lambda)}
+\frac{D_2D_3}{\lambda}.
\label{eq:5.6}
\end{equation}
As \(\lambda\to0\), \(D_2\to2D_1\), making \(D_1D_3\) and \(D_2D_3\) linearly dependent. The two terms with divergent coefficients in Eq.~\eqref{eq:5.6} must be combined before taking the limit:
\begin{equation}
\begin{aligned}
-\frac{4D_1D_3}{\lambda(2-\lambda)}
+\frac{D_2D_3}{\lambda}
&=
D_3\left[
-\frac{4x}{\lambda(2-\lambda)}
+\frac{2x+\lambda}{\lambda}
\right]
\\
&=
D_3\left(1-\frac{2x}{2-\lambda}\right)
\longrightarrow 1-x^2.
\end{aligned}
\label{eq:5.7}
\end{equation}
The remaining term and the combined pair have the limits
\begin{equation}
\frac{D_1D_2}{2-\lambda}
\longrightarrow x^2,
\qquad
x^2+(1-x^2)=1.
\label{eq:5.8}
\end{equation}
The divergences of \(c_{1,3}\) and \(c_{2,3}\) thus reflect the degeneration of the denominator-product basis. At fixed \(x\), combining the relevant terms preserves a finite polynomial identity in Eq.~\eqref{eq:5.6}. At \(\lambda=0\), the linear relation \(D_2=2D_1\) implies \(D_1D_2=2D_1^2\), and the three-term expansion valid at generic kinematics no longer has a finite termwise representation. A rational-function reduction performed directly at the special point must retain the corresponding repeated-propagator structure. The finiteness discussed here concerns only this algebraic limit and makes no assertion about the kinematic limit after integration.

\Needspace{10\baselineskip}
\section{Conclusions}\label{sec:conclusion}

We have given two complementary proofs of the determinantal coefficient formula introduced by Bargie\l a and Yang~\cite{BargielaYang2024} for high-multiplicity integrand reduction in the tHV scheme. The affine relations among propagators reduce the question to a rational-function identity with coefficients independent of loop momenta. This isolates the coefficient problem underlying the reduction to finite basis topologies and connects it to the earlier studies of denominator reducibility~\cite{GluzaKajdaKosower2011,Kleiss2012,Pak2012}.

The recursive proof explains how the compact expression arises from successive denominator removal. For three affine relations, the six ordered paths can be grouped according to their first step. Induction extends this organization to all \(A!\) removal orders: the unreplaced and replacement determinants of each smaller system acquire the same elimination factor, which cancels in the coefficient, while a cofactor identity combines the possible first steps. The resulting expression is independent of the chosen basis of affine relations. Thus the determinant formula summarizes the complete path sum, even though individual intermediate factors depend on the elimination procedure.

The cut-evaluation proof gives a direct interpretation of the same result. After clearing the common denominator, setting the \(r=m-A\) complementary denominators to zero leaves one nonzero product. The diagonal evaluation matrix and the dimension count establish the existence and uniqueness of the expansion, and Cramer's rule gives the full coefficient. This connects the proof to one-loop coefficient-extraction methods based on integrand decompositions and cut data~\cite{BrittoCachazoFeng2005,OPP2007,EllisGieleKunszt2008,MastroliaMirabellaPeraro2012}, and more specifically to residue extraction in denominator coordinates~\cite{FekeshazyKardos2026}. The complementary roles of the proofs are therefore clear: recursion explains how the determinant structure is assembled, while evaluation explains why each coefficient has its reciprocal-product form.

The discussion of special kinematics identifies the limits of this representation. A vanishing unreplaced determinant alone does not force a pole; the replacement determinants must also be examined. When the denominator-product basis degenerates, individual coefficients may diverge even though their combined polynomial identity has a finite limit. If the dimension of the relation space changes, the reduction must be reconstructed. These statements concern the algebraic identity before integration and do not determine the behavior of an integrated amplitude at the same kinematic point.

The present derivation provides an explicit reference for the unit-power, constant-coefficient case. Repeated propagators and more general numerator decompositions are natural directions for extending the determinant analysis, with existing polynomial-division and residue methods providing relevant starting points~\cite{MastroliaMirabellaPeraro2012,Mastrolia2013,FekeshazyKardos2026}. More general existence criteria for partial fractions on hyperplane arrangements~\cite{deKorteYu2026} may also help organize families for which the denominator-product basis used here is unavailable. Such extensions would require a separate treatment of the changed basis and coefficient structure.

\begin{acknowledgments}
I would like to thank Bo Feng and Tong-Zhi Yang for helpful discussions.
\end{acknowledgments}

\makeatletter
\immediate\write\@auxout{\string\citation{manuscriptControl}}
\makeatother
\bibliographystyle{apsrev4-2}
\bibliography{references}

\end{document}